\documentclass[12pt, a4paper]{article}
\usepackage{graphicx}
\usepackage{amssymb}
\usepackage{amsmath}
\usepackage{ascmac}
\usepackage{bm}
\usepackage{braket}
\usepackage[dvipsnames]{xcolor}
\usepackage{theorem}
\usepackage{subcaption}
\usepackage{listings}
\usepackage{colortbl}
\usepackage{tabularx}
\usepackage{longtable}
\usepackage[utf8]{inputenc}
\usepackage[T1]{fontenc}
\usepackage{lmodern}
\usepackage[top=30truemm,bottom=30truemm,left=25truemm,right=25truemm]{geometry}
\usepackage[sort&compress,numbers, merge]{natbib}
\usepackage{multirow}
\usepackage{slashed}
\usepackage{here}

\definecolor{Orange}{cmyk}{0,0.61,0.87,0}
\definecolor{JungleGreen}{cmyk}{0.99,0,0.52,0}
\definecolor{OliveGreen}{cmyk}{0.64,0,0.95,0.40}
\definecolor{Brown}{cmyk}{0,0.81,1,0.60}
\definecolor{RoyalBlue}{cmyk}{0.71,0.53,0,0.12}
\definecolor{Gray}{cmyk}{0,0,0,0.40}
\definecolor{LightPink}{cmyk}{0.0,0.25,0,0}
\definecolor{LLightPink}{cmyk}{0.0,0.10,0,0}
\definecolor{LightBlue}{cmyk}{0.25,0,0,0}
\definecolor{LightGray}{cmyk}{0,0,0,0.2}

\newcommand{\del}{\partial}

\newcommand{\nn}{\nonumber}

\allowdisplaybreaks[1]

\usepackage[colorlinks=true, linkcolor=OliveGreen, citecolor=RoyalBlue,
urlcolor=RoyalBlue]{hyperref}

\renewcommand{\thefootnote}{\fnsymbol{footnote}}

\begin{document}

\renewcommand{\thefootnote}{\fnsymbol{footnote}}

\begin{flushright}
{\tt 
KYUSHU-HET-361
\\
UT-HET-146
\\
KANAZAWA-26-03
}
\end{flushright}

\begin{center}
\textbf{\Large 
A radiative neutrino mass model with semi-annihilating boosted dark matter
}
\vspace{0.5cm}

Motoko Fujiwara$^{a,b}$,\footnote{E-mail address: \href{mailto:fujiwara.motoko@phys.kyushu-u.ac.jp}{\tt fujiwara.motoko@phys.kyushu-u.ac.jp}}
Takashi Toma$^{c,d}$\footnote{E-mail address: \href{mailto:toma@staff.kanazawa-u.ac.jp}{\tt toma@staff.kanazawa-u.ac.jp}}

\vspace{0.2cm}

{\small  \it$^a$ Department of Physics, Kyushu University, 744 Motooka, Nishi-ku, Fukuoka, 819-0395, Japan
}
\\
{\small  \it$^b$ Department of Physics, University of Toyama, 3190 Gofuku, Toyama 930-8555, Japan}
\\
{\small  \it$^c$ 
Institute of Liberal Arts and Science, Kanazawa University, Kanazawa 920-1192, Japan
}
\\
{\small  \it$^d$ 
Institute for Theoretical Physics, Kanazawa University, Kanazawa 920-1192, Japan
}

\vspace{0.25cm}

\today

\vspace{0.5cm}

\abstract{
Boosted dark matter can provide a distinctive signature of a non-minimal dark sector. 
In this work, we construct a radiative neutrino mass model that allows for the production of boosted dark matter in the present universe through semi-annihilation. 
A Dirac fermion serves as the dark matter candidate, which semi-annihilates into an anti-dark matter particle and a neutrino. Taking into account the relevant experimental and theoretical constraints, 
we find that the boosted dark matter can be probed by DARWIN if its mass lies in the range of 20–50 MeV and the semi-annihilation proceeds through a resonance with a tuning of $\mathcal{O}(0.1)\%$. 
Furthermore, the monochromatic neutrino produced simultaneously with the boosted dark matter can also be detected at Hyper-Kamiokande and DUNE in part of the allowed parameter space.
}

\end{center}

\setcounter{footnote}{0}
\renewcommand{\thefootnote}{\arabic{footnote}}

\newpage 
\section{Introduction}
The standard thermal freeze-out scenario based on $2\to2$ annihilation process is one of the most well-motivated mechanisms for producing dark matter, 
predicting a thermal dark matter mass in the range of $\mathcal{O}(1)~\mathrm{MeV}\sim\mathcal{O}(100)~\mathrm{TeV}$. 
Such dark matter candidates can be explored by various experiments including direct detection, indirect detection, collider searches, and cosmological observations. 
However, no confirmed signal has been observed in the parameter regions probed so far, leading to strong constraints in several well-motivated regions of parameter space. 
Consequently, alternative dark matter candidates such as axions and primordial black holes have attracted increasing attention.

On the other hand, a non-minimal extension of the thermal dark matter scenario may evade the existing constraints and opens up new possibilities for thermal dark matter. 
In this case, non-standard dark matter annihilation processes of dark matter may appear, such as semi-annihilations $\chi\chi\to\overline{\chi}\phi$~\cite{Hambye:2008bq, DEramo:2010keq}, 
Strongly Interacting Massive Particle (SIMP) processes including $\chi\chi\chi\to\chi\overline{\chi}$ and $\chi\chi\chi\chi\to\chi\chi$~\cite{Hochberg:2014dra}, 
and dark matter conversion processes $\chi_2\chi_2\to\chi_1\chi_1$ in multi-component dark matter models. 
Here $\chi~(\overline{\chi})$ denotes the (anti-)dark matter particle, $\phi$ is either a Standard Model (SM) or a new light particle, 
and $\chi_{2}~(\chi_1)$ represents the heavier (lighter) dark matter component~\cite{Belanger:2011ww, Agashe:2014yua}. 
In these scenarios, these processes play a central role in determining the thermal relic abundance and can also produce energetic dark matter particles in the final state.

An distinctive feature of non-minimal extensions is that some dark matter particles are produced in the final state with kinematically determined energies, in the present universe,
since the dark matter particles in the initial state are non-relativistic. 
Such boosted dark matter does not arise in the minimal thermal dark matter scenario and can be a distinctive signature of a non-minimal dark sector. 
It may be detected through dark matter direct detection experiments~\cite{Cherry:2015oca, Giudice:2017zke} 
and large-volume neutrino experiments~\cite{Agashe:2014yua} (see also~\cite{Kim:2016zjx, Toma:2021vlw, Aoki:2023tlb, BetancourtKamenetskaia:2025noa, BetancourtKamenetskaia:2025ivl, BetancourtKamenetskaia:2026tfx} and comprehensive studies for boosted dark matter~\cite{Berger:2022cab, Choi:2025wbw}). 
In particular, previous work~\cite{BetancourtKamenetskaia:2025noa} found that a scattering cross section larger than $\mathcal{O}(10^{-38})~\mathrm{cm^2}$ is needed to detect boosted dark matter produced by the semi-annihilation $\chi\chi\to\overline{\chi}\nu$ for sub-GeV dark matter masses where $\nu$ denotes an active neutrino. 
However, realizing such a large cross section while satisfying the other experimental and theoretical constraints 
such as electroweak precision tests, the observed dark matter relic abundance and indirect detection of dark matteris highly non-trivial. 
This motivates the construction of a concrete model.\footnote{The semi-annihilation is also possible in the radiative neutrino mass models proposed in Refs.~\cite{Ma:2007gq, Aoki:2014cja}. However, the dark matter candidate in those models is leptophilic and therefore does not scatter off protons, making boosted dark matter detection via proton scattering highly suppressed.}

In this work, we construct a concrete model that produces boosted dark matter from semi-annihilations. 
In our model, we add several Dirac fermions and complex scalars to the Standard Model, and impose a global U(1)$_L$ symmetry, which is spontaneously broken to a residual $\mathbb{Z}_3$ symmetry. 
The residual $\mathbb{Z}_3$ symmetry plays an important role in forbidding tree-level neutrino masses and ensuring stability of the lightest neutral particle, which serves as the dark matter candidate at the same time, 
while the small neutrino masses are generated by two-loop diagrams, providing a natural explanation for their smallness. 
As a result, our model establishes a close connection between dark matter and neutrino phenomenology.
In particular, the semi-annihilation into an anti-dark matter particle and a neutrino is allowed by the residual $\mathbb{Z}_3$ symmetry, realizing boosted dark matter signatures.
Previous studies~\cite{BetancourtKamenetskaia:2025noa, BetancourtKamenetskaia:2025ivl} have investigated the model-independent studies of boosted dark matter detection. 
We incorporate these results into a neutrino mass model framework, and study the detection prospects of boosted dark matter as well as the accompanying neutrinos 
through dark matter direct detection experiment DAWRIN and neutrino experiment DUNE.

This paper is organized as follows. 
In Section~\ref{sec:model}, we present a concrete realization of a boosted dark matter model with a global U(1)$_L$ symmetry and derive the mass spectrum. 
In Section~\ref{sec:nu_masses}, we present the formulae for the small neutrino masses induced at the two-loop level and discuss their correlation with dark matter phenomenology. 
In Section~\ref{sec:const}, we discuss the relevant experimental and theoretical constraints for subsequent numerical calculations. 
In Section~\ref{sec:dark}, we identify the lightest Dirac fermion as the dark matter candidate and study its phenomenology. 
The thermal relic abundance is determined by the semi-annihilation or the standard annihilation process, depending on the parameter region. 
Some parameter regions predict both boosted dark matter and monochromatic neutrino signals that can be probed by future direct detection and neutrino experiments at the same time.
Finally, our conclusions are given in Section~\ref{sec:conc}.

\section{Model with a global U(1)$_L$ symmetry}
\label{sec:model}

\subsection{Lagrangian}

We consider an extension of the two-loop radiative neutrino mass model~\cite{Ma:2007gq} with a global U(1)$_L$ symmetry, which is identified 
with the lepton number.\footnote{One may also consider an extension with a gauge symmetry instead of a global symmetry though it is beyond the scope of this work.} 
The particle contents are shown in Table~\ref{tab:particles_real-singlet}. 
The singlet scalar $\Sigma$ is introduced to spontaneously break the U(1)$_L$ symmetry down to the residual $\mathbb{Z}_3$ symmetry 
and serves as the mediator of elastic scattering between dark matter and protons for direct detection. 
To determine a consistent charge assignment, we follow the classification obtained for the gauged U(1)$_{B-L}$ symmetry in Ref.~\cite{Ho:2016aye}, 
and we adopt the same assignment (C8) in Table VI that reference as the simplest option. 
The other assignments (C1), (C3), (C4), (C5), and (C6) in Ref.~\cite{Ho:2016aye} also lead to qualitatively similar results.

\begin{table}[tb]
\centering
\begin{tabular}{c||c|c|c|c||c|c|c|c}
  &  ~~$L_L$~~  &  ~~$e^c_R$~~  &  ~~$\psi_L$~~  &  ~~$\psi_R^c$~~ & ~~$H$~~ & ~~$\Sigma$~~ & ~~$\eta$~~ & ~~$\chi$~~
\\
\hline
\hline
SU(2)$_L$  &  $\bm{2}$  &  $\bm{1}$  &  $\bm{1}$  &  $\bm{1}$ & $\bm{2}$ & $\bm{1}$ & $\bm{2}$ & $\bm{1}$
\\
\hline
U(1)$_{Y}$  &  $-1/2$  &  $1$  &  $0$  &  $0$ & $1/2$ & $0$ & $1/2$ & $0$
\\
\hline
U(1)$_L$  &  $1$  &  $-1$  &  $-1/3$  &  $-5/3$ & $0$ & $2$ & $2/3$ & $2/3$
\\
\hline
Residual $\mathbb{Z}_3$  &  $0$  &  $0$  &  $1$  &  $-1$ & $0$ & $0$ & $1$ & $1$
\\
\hline
Spin & $1/2$ & $1/2$ & $1/2$ & $1/2$ & $0$ & $0$ & $0$ & $0$
\end{tabular}
\caption{Particle contents and charge assignments for fermions and scalars where the flavor indices are omitted. 
}
\label{tab:particles_real-singlet}
\end{table}

The Beyond SM (BSM) Lagrangian is given by
\begin{align}
{\cal  L}_{\rm  BSM}  &=  
\sum_i  \overline{\psi}_i i \slashed{\del} \psi_i
+  \left|\del_\mu \Sigma\right|^2
+  \left| D_\mu  \eta \right|^2
+  \left|\del_\mu  \chi\right|^2
\nn
\\
&\quad
-  \bigg( 
   y_{\alpha}  H^\dagger  \overline{e}_\alpha  L_\alpha
+  (y_{\nu})_{i\alpha}  \eta  \overline{\psi}_i  P_L  L_\alpha  
+  \frac{(y_L)_{ij}}{2}  \chi  \overline{\psi_i^c}  P_L  \psi_j
+  (y_\Sigma)_{i} \Sigma  \overline{\psi}_i P_L \psi_i
+  \text{h.c.}
\bigg),
\end{align}
where $\alpha = e,\mu,\tau$ and $i,j=1,2,3$ denote the flavor and generation indices, respectively. 
The charged lepton Yukawa couplings $y_\alpha$ and the Dirac Yukawa coupling $y_{\Sigma}$ can be diagonalized without loss of generality. 
As we will see later, the complex scalar $\Sigma$ acquires a vacuum expectation value (VEV). 
The Dirac fermion $\psi_i$ obtains a mass $m_{i}\equiv (y_{\Sigma})_i \langle \Sigma\rangle$ via the Yukawa coupling,  
and the lightest $\psi_i$ can be a dark matter candidate. 
As discussed below, the Yukawa coupling $y_L$ controls the interaction between $\chi$ and $\psi_i$, 
and plays a crucial role in the detection of this dark matter candidate via the semi-annihilation.

The most general scalar potential is given by
\begin{align}
V( H, \eta, \chi, \sigma )
&=
\mu_H^2 |H|^2  
+  \mu^2_\Sigma  |\Sigma|^2
+  \mu_\eta^2 |\eta|^2
+  \mu_\chi^2 |\chi|^2
+  \frac{\lambda_H}{4} |H|^4
+  \frac{\lambda_{\Sigma}}{4}  |\Sigma|^4
+  \frac{\lambda_\eta}{4} |\eta|^4
+  \frac{\lambda_{\chi}}{4} |\chi|^4
\nn
\\
&  \quad
+  \lambda_{H  \Sigma} |H|^2 |\Sigma|^2
+  \lambda_{H\eta} |H|^2 |\eta|^2
+  \lambda_{H\eta}'  ( H^\dagger  \eta )  ( \eta^\dagger  H )
+  \lambda_{H  \chi} |H|^2 |\chi|^2 
\nn
\\
&  \quad
+  \lambda_{\Sigma  \eta} |\Sigma|^2 |\eta|^2
+  \lambda_{\Sigma  \chi} |\Sigma|^2 |\chi|^2
+  \lambda_{\eta  \chi} |\eta|^2 |\chi|^2 
\nn
\\
&  \quad
+  
\left( 
\tilde{\mu}_{\chi}  ( H^\dagger  \eta )  \chi^\dagger
+  \frac{\tilde{\lambda}_{\chi}}{3!}  \Sigma^\dag  \chi^3
+  \text{h.c.} \right),
\label{eq:potential}
\end{align}
where the couplings $\tilde{\mu}_\chi$ and $\tilde{\lambda}_\chi$ are generically complex; however, their complex phases can be absorbed by field redefinitions, 
and thus the scalar potential is CP conserving. 
Consequently, the scalar potential contains $17$ independent real parameters.

\subsection{Scalar sector}

We derive the mass eigenstates by diagonalizing the scalar mass matrix after assuming the following VEVs:
\begin{align}
  \Braket{H}
  &=  
  \begin{pmatrix}
  0
  \\
  v/\sqrt{2}
  \end{pmatrix},
  &
  \Braket{\eta}
  &=
  \begin{pmatrix}
  0
  \\
  0
  \end{pmatrix},
  &
  \Braket{\chi}  &=  0,
  &
  \Braket{\Sigma}  &=  \frac{v_\sigma}{\sqrt{2}}.
  &
  \label{eq:vac}
\end{align}
Expanding the scalar fields around the VEVs, we write
\begin{align}
  H
  &=  
  \begin{pmatrix}
  \pi^+
  \\
  \frac{v  + h_\text{SM} +  i  \pi_Z}{\sqrt{2}}
  \end{pmatrix},
  &
  \eta
  &=
  \begin{pmatrix}
  \eta^+
  \\
  \eta^0
  \end{pmatrix}
  &
  \chi  &=  \chi,
  &
  \Sigma  &=  \frac{v_\sigma  +  \sigma + iJ}{\sqrt{2}}.
  &
  \label{eq:scalars}
\end{align}
The fields $\pi^\pm$ and $\pi_Z$ are would-be Nambu-Goldstone bosons for $W^\pm$ and $Z$, respectively. 
The stationary conditions to realize the desired VEVs are given by
\begin{align}
  \mu_H^2 + \frac{\lambda_H}{4} v^2  +  \frac{\lambda_{H  \Sigma}}{2}  v_\sigma^2 = &~0,
  \\
  \mu_\Sigma^2 + \frac{\lambda_{H  \Sigma}}{2}    v^2  +  \frac{\lambda_\Sigma}{4}  v_\sigma^2 = &~0. 
\end{align}
Using these conditions, the mass matrix for the CP-even scalars is obtained as
\begin{align}
 V\supset \frac{1}{2}\left(
\begin{array}{cc}
 \sigma & h_\mathrm{SM}
\end{array}
\right)\left(
\begin{array}{cc}
 \frac{1}{2}\lambda_{\Sigma}v_\sigma^2 & \lambda_{H\Sigma}vv_\sigma\\
\lambda_{H\Sigma}vv_\sigma & \frac{1}{2}\lambda_H v^2
\end{array}
\right)\left(
\begin{array}{c}
 \sigma\\
h_\mathrm{SM}
\end{array}
\right),
\end{align}
and is diagonalized by the mass eigenstates $s$ and $h$ (the SM-like Higgs boson with the mass $m_h=125~\mathrm{GeV}$) as
\begin{align}
\left(
\begin{array}{c}
 \sigma\\
 h_\text{SM}
\end{array}
\right)=\left(
\begin{array}{cc}
 \cos\theta & -\sin\theta\\
\sin\theta & \cos\theta
\end{array}
\right)\left(
\begin{array}{c}
 s\\
 h
\end{array}
\right),
\end{align}
where the mixing angle is given by
\begin{align}
\tan2\theta=\frac{2\lambda_{H\Sigma}vv_\sigma}{\frac{1}{2}\lambda_{\Sigma}v_\sigma^2 - \frac{1}{2}\lambda_{H}v^2}.
\end{align}

Similarly, the mass matrix for the $\mathbb{Z}_3$-charged scalars $\eta^0$ and $\chi$ is also derived as 
\begin{align}
V\supset \left(
\begin{array}{cc}
 \chi^\dag & \eta^{0\dag}\\
\end{array}
\right)\left(
\begin{array}{cc}
 \mu_\chi^2+\frac{1}{2}\lambda_{H\chi}v^2+\frac{1}{2}\lambda_{\Sigma\chi}v_\sigma^2 & \frac{1}{\sqrt{2}}\tilde{\mu}_{\chi}v\\
\frac{1}{\sqrt{2}}\tilde{\mu}_{\chi}v & \mu_\eta^2+\frac{1}{2}\left(\lambda_{H\eta}+\lambda_{H\eta}^\prime\right)v^2+\frac{1}{2}\lambda_{\Sigma\eta}v_\sigma^2
\end{array}
\right)\left(
\begin{array}{c}
 \chi\\
\eta^0
\end{array}
\right).
\end{align}
The mass matrix can be diagonalized and the gauge eigenstates are related to the mass eigenstates $\varphi$ and $\tilde{\varphi}$ as
\begin{align}
 \left(
\begin{array}{c}
 \chi\\
 \eta^0
\end{array}
\right)=\left(
\begin{array}{cc}
\cos\alpha & -\sin\alpha\\
\sin\alpha & \cos\alpha
\end{array}
\right)\left(
\begin{array}{c}
 \varphi\\
\tilde{\varphi}
\end{array}
\right),
\end{align}
where the mixing angle is given by
\begin{align}
\tan2\alpha=\frac{\sqrt{2}\tilde{\mu}_{\chi}v}
{\mu_\chi^2-\mu_\eta^2+\frac{1}{2}\left(\lambda_{H\chi}-\lambda_{H\eta}-\lambda_{H\eta}^\prime\right)v^2+\frac{1}{2}\left(\lambda_{\Sigma\chi}-\lambda_{\Sigma\eta}\right)v_\sigma^2}.
\end{align}
The mass of the charged scalar $\eta^\pm$ is given by $m_\eta^2=\mu_\eta^2+\frac{\lambda_{H\eta}}{2}v^2+\frac{\lambda_{\Sigma\eta}}{2}v_\sigma^2$. 

The vacuum in Eq.~(\ref{eq:vac}) preserves the residual $\mathbb{Z}_{3}$ symmetry, which ensures the stability of the dark matter candidate $\psi_1$. 
The conditions for the vacuum in Eq.~(\ref{eq:vac}) to be a local minimum are that the mass matrices for the CP-even scalars and $\mathbb{Z}_3$-charged scalars are positive definiteness. 
These conditions are explicitly given by
\begin{align}
&\hspace{1cm}\lambda_{\Sigma} > 0,\quad
\lambda_{H} > 0,\quad
\lambda_{\Sigma}\lambda_{\Sigma} > 4\lambda_{H\Sigma}^2,\quad
m_{\eta}^2 > 0,\nonumber\\
&\mu_{\chi}^2+\frac{\lambda_{H\chi}}{2}v^2+\frac{\lambda_{\Sigma\chi}}{2}v_\sigma^2 > 0,\quad
\mu_{\eta}^2+\frac{\lambda_{H\eta}+\lambda_{H\eta}^\prime}{2}v^2+\frac{\lambda_{\Sigma\eta}}{2}v_\sigma^2 > 0,\nonumber\\
&\left(\mu_{\chi}^2+\frac{\lambda_{H\chi}}{2}v^2+\frac{\lambda_{\Sigma\chi}}{2}v_\sigma^2\right) 
\left(\mu_{\eta}^2+\frac{\lambda_{H\eta}+\lambda_{H\eta}^\prime}{2}v^2+\frac{\lambda_{\Sigma\eta}}{2}v_\sigma^2\right) > \frac{\tilde{\mu}_\chi^2v^2}{2}.
\end{align}
A deeper minimum may exist if $\tilde{\lambda}_\chi$ in Eq.~(\ref{eq:potential}) is large enough. 
Even in this case, the local minimum remains phenomenologically acceptable if the vacuum decay rate into the global minimum is longer than the age of the universe. 

The CP-odd component of $\Sigma$, denoted by $J$ is the Nambu-Goldstone boson associated with the global U(1)$_L$ symmetry. 
Its mass can arise from higher-dimensional operators involving an additional field that acquires a large VEV~\cite{Ibarra:2013eda}, 
and allowing $J$ to be much heavier than the electroweak scale. 
In this work, we assume that $J$ is much heavier than the dark matter and is therefore decoupled.

\section{Small neutrino masses}
\label{sec:nu_masses}

\begin{figure}[t]
 \begin{center}
 \includegraphics[width=7cm]{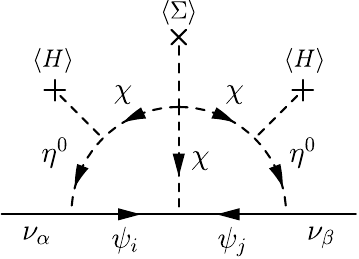}
\caption{Feynman diagram for small neutrino masses at two loop level.}
\label{fig:diagram}
 \end{center}
\end{figure}

The active neutrino masses are generated by the two-loop diagram shown in Fig.~\ref{fig:diagram} and are given by~\cite{Aoki:2014cja}
\begin{align}
\left(m_{\nu}\right)_{\alpha\beta}=&~
\frac{\tilde{\lambda}_\chi v_\sigma \sin^22\alpha}{4\sqrt{2}(4\pi)^4}
\sum_{i,j}
(y_{\nu})_{i\alpha} (y_{L})_{ij} (y_{\nu})_{j\beta}\nonumber\\
&\times\Big[
\cos^2\alpha \left\{
I\left(\xi_{i}^{\varphi},\xi_{ij}^{\varphi\varphi}\right)
-I\left(\xi_{i}^{\varphi},\xi_{ij}^{\varphi\tilde{\varphi}}\right)
-I\left(\xi_{i}^{\tilde{\varphi}},\xi_{ij}^{\varphi\varphi}\right)
+I\left(\xi_{i}^{\tilde{\varphi}},\xi_{ij}^{\varphi\tilde{\varphi}}\right)
\right\}\nonumber\\
&\hspace{0.7cm}
+\sin^2\alpha \left\{
I\left(\xi_{i}^{\tilde{\varphi}},\xi_{ij}^{\tilde{\varphi}\tilde{\varphi}}\right)
-I\left(\xi_{i}^{\tilde{\varphi}},\xi_{ij}^{\tilde{\varphi}\varphi}\right)
-I\left(\xi_{i}^{\varphi},\xi_{ij}^{\tilde{\varphi}\tilde{\varphi}}\right)
+I\left(\xi_{i}^{\varphi},\xi_{ij}^{\tilde{\varphi}\varphi}\right)
\right\}
\Big]
\nonumber\\
\simeq&~
\frac{\tilde{\lambda}_\chi v_\sigma \sin^22\alpha\cos^2\alpha}{4\sqrt{2}(4\pi)^4}
\sum_{i,j}
(y_{\nu})_{i\alpha}(y_L)_{ij}(y_{\nu})_{j\beta}I\left(\xi_{i}^{\varphi},\xi_{ij}^{\varphi\varphi}\right),
\label{eq:numass}
\end{align}
where we have used $m_{\varphi}^2\ll m_{\tilde{\varphi}}^2$ in the last step. 
The dimensionless parameters $\xi_{i}^a$ and $\xi_{ij}^{bc}$ are defined by~\cite{Aoki:2014cja}
\begin{align}
\xi_{i}^{a}\equiv\frac{m_a^2}{m_i^2},\quad
\xi_{ij}^{bc}\equiv\frac{xm_j^2+ym_b^2+zm_c^2}{y(1-y)m_i^2},
\end{align}
and the loop function is given by
\begin{align}
 I\left(\xi_{i}^{a},\xi_{ij}^{bc}\right)\equiv
\frac{m_j}{m_i}
\int_{0}^{1}dxdydz \frac{\delta\left(1-x-y-z\right)}{y(1-y)}
\frac{1}{\xi_i^{a}-\xi_{ij}^{bc}}
\left[
\frac{\xi_{i}^{a}\log\xi_{i}^a}{1-\xi_{i}^a}
-\frac{\xi_{ij}^{bc}\log\xi_{ij}^{bc}}{1-\xi_{ij}^{bc}}
\right],
\end{align}
where $x,y$ and $z$ denote the Feynman parameters. 
Note that the loop function is symmetric under the simultaneous exchange of $i\leftrightarrow j$ and $a\leftrightarrow c$, 
namely $I\left(\xi_i^a,\xi_{ij}^{bc}\right)=I\left(\xi_j^c,\xi_{ji}^{ba}\right)$ though it is not manifest from the above expression.\footnote{This symmetry becomes manifest in the original momentum-integral representation before introducing the Feynman parametrization.} 

One can solve Eq.~(\ref{eq:numass}) in terms of $(\tilde{y}_L)_{ij}\equiv (y_L)_{ij}I(\xi_{i}^{\varphi},\xi_{ij}^{\varphi\varphi})$ as 
\begin{align}
\tilde{y}_L=\frac{4\sqrt{2}(4\pi)^4}{\tilde{\lambda}_{\chi}v_\sigma \sin^22\alpha\cos^2\alpha}{y_{\nu}^{T}}^{-1} m_{\nu} y_{\nu}^{-1}.
\end{align}
Therefore, assuming a diagonal neutrino Yukawa coupling $y_{\nu}$, 
which is required to evade the strong constraints from the charged lepton flavor violating processes, 
we obtain a correlation between the effective mass for neutrinoless double beta decay $(m_{\nu})_{ee}$ and the Yukawa coupling $y_L$ relevant for dark matter phenomenology as
\begin{align}
(y_L)_{11}&=\frac{4\sqrt{2}(4\pi)^4(m_{\nu})_{ee}}{\tilde{\lambda}_\chi v_\sigma \sin^22\alpha\cos^2\alpha~ (y_{\nu})_{1e}^2 I\left(\xi_{1}^{\varphi},\xi_{11}^{\varphi\varphi}\right)}\nonumber\\
 &\approx0.2\left(\frac{0.01}{(y_{\nu})_{1e}\sin2\alpha}\right)^2\left(\frac{1~\mathrm{GeV}}{\tilde{\lambda}_\chi v_{\sigma}}\right)
 \left(\frac{0.1}{I(\xi_1^{\varphi},\xi_{11}^{\varphi\varphi})}\right)\left(\frac{(m_{\nu})_{ee}}{10^{-2}~\mathrm{eV}}\right),
\label{eq:yukawa}
\end{align}
where we have used $\sin\alpha\ll1$. 
Note that even if the neutrino Yukawa coupling $y_\nu$ is diagonal, the observed neutrino mixing angles can be reproduced through the Yukawa coupling $y_L$. 
The magnitude of the neutrino mass parameter $(m_\nu)_{ee}$ depends on the lightest neutrino mass~\cite{Agostini:2022zub}. 
In our model, the $(y_L)_{11}$ plays a crucial role in determining the thermal relic abundance of dark matter as we will study later, and is closely correlated with $(m_\nu)_{ee}$ via Eq.~(\ref{eq:yukawa}). 
Consequently, $(m_\nu)_{ee}$ cannot be as large as sub-eV values required for the observation of neutrinoless double beta decay. 
Throughout this work, we take the effective neutrino mass parameter to be $(m_\nu)_{ee}=10^{-5}~\mathrm{meV}$ for studying dark matter phenomenology. 
This choice implies that the lightest neutrino mass eigenvalue is constrained to lie in the range~\cite{Agostini:2022zub}
\begin{align}
2.1\times10^{-3}~\mathrm{eV} \lesssim (m_{\nu})_\mathrm{lightest} \lesssim 6.7\times10^{-3}~\mathrm{eV}\quad \text{for normal ordering},
\end{align}
and inverted ordering is inconsistent with the dark matter phenomenology discussed below.

\section{Constraints} 
\label{sec:const}

\subsection{Charged lepton flavor violation}
The charged lepton flavor violating processes (cLFV), $\ell_\alpha\to \ell_\beta\gamma$, place stringent constraints on the model. 
In particular, the most stringent constraint comes from $\mu\to e\gamma$, $\mathrm{Br}\left(\mu\to e\gamma\right)\leq 1.5\times10^{-13}$ at the $90\%$ confidence level~\cite{MEGII:2025gzr}. 
The decay width for the cLFV process $\ell_\alpha\to \ell_\beta\gamma$ in this model is given by~\cite{Ma:2001mr, Toma:2013zsa}
\begin{align}
\text{Br}(\ell_\alpha\to\ell_\beta\gamma)=
\frac{3\alpha_\mathrm{em}}{64\pi G_F^2m_\eta^4}\left|\sum_{i=1}^{3}(y_{\nu})_{i\alpha}(y_{\nu})_{i\beta}^* F_2\left(\frac{m_i^2}{m_{\eta}^2}\right)\right|^2
\mathrm{Br}\left(\ell_\alpha\to\ell_\beta\nu_\alpha\overline{\nu}_\beta\right),
\end{align}
where $G_F$ is the Fermi constant and the loop function $F_2(x)$ is given by
\begin{align}
F_2(x)=\frac{1-6x+3x^2+2x^3-6x^2\log{x}}{6(1-x)^4}.
\end{align}
The above expression for the branching ratio shows that this strong bound can be avoided by choosing a diagonal neutrino Yukawa coupling $y_{\nu}$, and this choice is adopted throughout this work. 
The neutrino mixing angles can then be reproduced by the flavor structure of the Yukawa coupling $y_L$.

\subsection{Electroweak precision test}
The new contribution to the $T$ parameter (equivalently, the $\rho$ parameter), is given by~\cite{Barbieri:2006dq, Aoki:2014cja}
\begin{align}
\Delta T=\frac{2}{(4\pi)^2 \alpha_\mathrm{em}v^2}\left[
\cos^2\alpha F\left(m_\eta^2,m_{\tilde{\varphi}}^2\right)
+\sin^2\alpha F\left(m_\eta^2,m_{\varphi}^2\right)
-\cos^2\alpha\sin^2\alpha F\left(m_{\tilde{\varphi}}^2,m_{\varphi}^2\right)
\right],
\end{align}
where the loop function $F(x,y)$ is given by
\begin{align}
F(x,y)=\frac{x+y}{2}-\frac{xy}{x-y}\log\left(\frac{x}{y}\right). 
\end{align}
For $m_{\varphi}\ll m_{\tilde{\varphi}}\approx m_\eta$ in our setup, $\Delta T$ reduces to
\begin{align}
\Delta T\approx
\frac{2m_\eta^2}{(4\pi)^2 \alpha_\mathrm{em} v^2}\left[
\cos^2\alpha\left(1-\frac{m_{\tilde{\varphi}}^2}{m_\eta^2}\right)^2 + \frac{\sin^4\alpha}{2}
\right].
\end{align}
The experimental bound is given by $\Delta T = 0.09 \pm 0.13$ at $1\sigma$ level~\cite{Baak:2014ora}. 
This leads to bounds on the mass splitting between $m_{\tilde{\varphi}}$ and $m_\eta$, and $\sin\alpha$ as 
\begin{align}
\left|1-\frac{m_{\tilde{\varphi}}}{m_\eta}\right|\lesssim0.014,\quad
\sin\alpha\lesssim0.20.
\end{align}

The light complex scalar $\varphi$ gives a new contribution to the invisible decay of the $Z$ boson ($Z\to \varphi^*\varphi$) through the mixing angle $\sin\alpha$ 
since the produced $\varphi$ subsequently decays into $\psi\nu_e$, which are invisible to the detector.
The corresponding decay width is given by
\begin{align}
\Gamma_\mathrm{inv}=\frac{G_F m_Z^3}{24\sqrt{2}\pi}\sin^4\alpha\left(1-\frac{4m_{\varphi}^2}{m_Z^2}\right)^{3/2}.
\end{align}
The invisible decay width is constrained to satisfy $\Gamma_\mathrm{inv}\leq 2.0~\mathrm{MeV}$ by the LEP experiment at the $95\%$ confidence level~\cite{ALEPH:2005ab}. 
This implies the bound on the mixing angle $\sin\alpha\lesssim0.39$.

\subsection{Higgs mixing angle}
\label{sec:const_Hmix}

The Higgs mixing angle $\sin\theta$ is constrained to lie in the range $10^{-5}\lesssim \sin\theta \lesssim 10^{-4}$ for $m_s\lesssim200~\mathrm{MeV}$~\cite{Goudzovski:2022vbt}, 
where the upper and lower bounds come from the $K^+\to \pi^++\mathrm{inv.}$ search (NA62)~\cite{NA62:2021zjw} and from Big Bang nucleosynthesis (BBN)~\cite{Planck:2018vyg}, respectively. 
Most of the parameter space would be excluded by the BBN bound if the mediator mass is as light as $m_s\lesssim 2m_e\approx 1~\mathrm{MeV}$, because the mediator decay $s\to e^+e^-$ is kinematically forbidden 
and it becomes long-lived. 
The decay width of the mediator $s$ into $e^+e^-$ for $m_s\geq 2m_e$ is calculated as
\begin{align}
\Gamma\left(s\to e^+e^-\right)=\frac{\sin^2\theta m_s m_e^2}{8\pi v^2}\left(1-\frac{4m_e^2}{m_s^2}\right)^{3/2}
\approx\frac{1}{0.038~\mathrm{s}}\left(\frac{m_s}{10~\mathrm{MeV}}\right)\left(\frac{\sin\theta}{10^{-4}}\right)^2.
\end{align}

\subsection{Potential stability}

The quartic coupling $\tilde{\lambda}_\chi$ required for generating the small neutrino masses contributes to the scattering amplitude for $\chi\chi\to\chi\chi$ at the one-loop level 
after the spontaneous breaking of U(1)$_L$ symmetry. 
The resulting loop-induced amplitude can be interpreted as an effective quartic coupling $\lambda_\chi$ in Eq.~(\ref{eq:potential}), 
and therefore the total quartic coupling, including both the tree-level and loop-induced contributions, must remain positive to ensure the stability of the scalar potential. 
This potential stability condition gives an upper bound on the coupling $\tilde{\lambda}_\chi$ as 
\begin{align}
\lambda_\chi-\frac{1}{96\sqrt{3}\pi}\left(\frac{\tilde{\lambda}_\chi v_\sigma}{m_\chi}\right)^4 \gtrsim 0,
\end{align}
which reduces to $\left|\tilde{\lambda}_\chi\right| v_\sigma/m_\chi \lesssim 4.8\lambda_\chi^{1/4}$. 
This constraint gives a lower bound on the Yukawa coupling $(y_L)_{11}$ via Eq.(\ref{eq:yukawa}), which plays a crucial role in the semi-annihilation of dark matter as discussed below.

\section{Dark matter}
\label{sec:dark}

We identify the lightest Dirac fermion $\psi_1$ as the dark matter candidate, hereafter denoted by $\psi$ with mass $m_\psi$. 
In this work, we are interested in the boosted dark matter produced by the semi-annihilations $\psi\psi\to\overline{\psi}\nu_\alpha$ 
and $\overline{\psi}\overline{\psi}\to\psi \nu_\alpha$. 
Although conventional direct detection experiments which assume a non-relativistic Galactic dark matter flux as an initial state lose sensitivity in the sub-GeV dark matter mass range, 
distinctive signatures of semi-annihilation boosted dark matter offer a unique probe of this mass scale~\cite{BetancourtKamenetskaia:2025noa}.

The parameters relevant to dark matter phenomenology are listed below.
\begin{align}
(y_{\nu})_{1e},
m_\psi, m_\varphi, m_s, \sin2\alpha, \sin\theta,
\end{align}
where $m_\psi=(y_\Sigma)_{1} v_\sigma/\sqrt{2}$. 
The remaining couplings are fixed to be $\tilde{\lambda}_\chi=0.1$ and $(y_\Sigma)_1=1.3$ throughout the following analysis.
The former controls the small neutrino mass scale, while the latter determines the dark matter mass.
Note that the Yukawa coupling $(y_L)_{11}$ is not an independent parameter through the small neutrino mass formula given by Eq.~(\ref{eq:yukawa}).

As a benchmark parameter set, we choose 
\begin{align}
\sin\theta=10^{-4},
\hspace{0.5cm}
1.4\lesssim m_s/m_\psi \lesssim 2,
\hspace{0.5cm}
m_{\tilde{\varphi}}\simeq m_{\eta}=3~\mathrm{TeV},
\label{eq:bp}
\end{align}
and the following mass hierarchy
\begin{align}
m_{\psi} \lesssim m_s \lesssim m_\varphi \ll m_h~(=125~\mathrm{GeV}) \ll m_{\tilde{\varphi}}\simeq m_\eta.
\end{align}
In Eq.~(\ref{eq:bp}), the Higgs mixing angle $\sin\theta$ is fixed close to its current upper bound~\cite{Goudzovski:2022vbt} (see the discussion in Sec.~\ref{sec:const_Hmix}).

\subsection{Relic abundance}

\begin{figure}[t]
 \begin{center}
 \includegraphics[width=14cm]{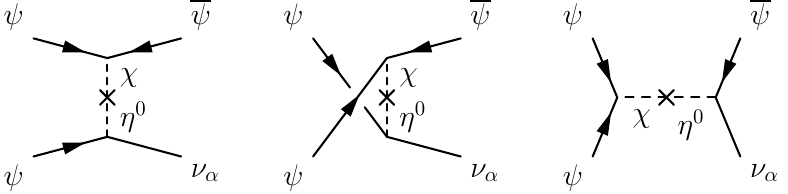}
\caption{Feynman diagrams for the semi-annihilation process $\psi\psi\to\overline{\psi}\nu_\alpha$.}
\label{fig:semi-ann}
 \end{center}
\end{figure}

\begin{figure}[t]
 \begin{center}
 \includegraphics[width=9cm]{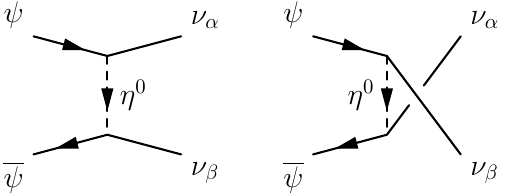}
\caption{Feynman diagrams for the standard annihilation process $\psi\overline{\psi}\to \nu_\alpha\nu_\beta$.}
\label{fig:ann}
 \end{center}
\end{figure}

\begin{figure}[t]
\begin{center}
\includegraphics[width=7.8cm]{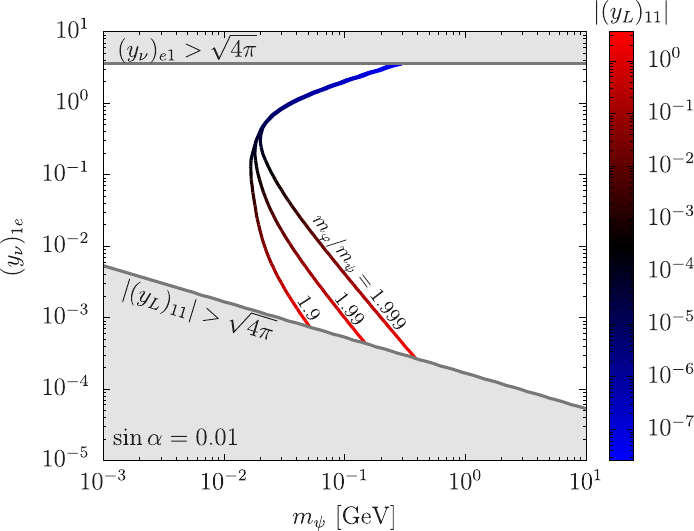}
\includegraphics[width=7.9cm]{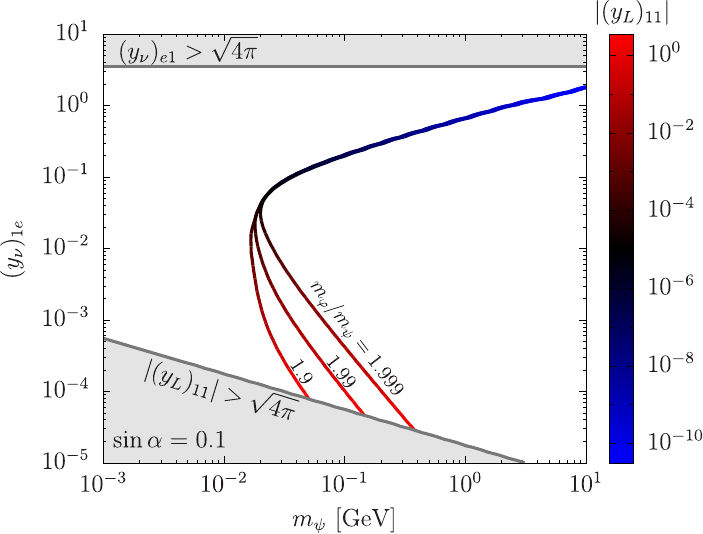}
\caption{
Parameter space reproducing the observed relic abundance $\Omega_\mathrm{DM} h^2=0.12$ where 
the mass of the $\mathbb{Z}_3$ charged scalar $\varphi$ is fixed to be $m_\varphi/m_\psi=1.9,1.99$ and $1.999$, and the mixing angle is $\sin\alpha=0.01$ (left) and $0.1$ (right).
The other parameters are set to be $\sin\theta=10^{-4}$ and $m_{\tilde{\varphi}}\approx m_\eta=3~\mathrm{TeV}$. 
}
\label{fig:relic}
\end{center}
\end{figure}

The thermal relic abundance of dark matter is required to reproduce  the observed value $\Omega_\mathrm{DM}h^2=0.12$~\cite{Planck:2018vyg}. 
In the parameter space we are interested in, the relic abundance is dominantly determined by the semi-annihilation 
$\psi\psi\to\overline{\psi}\nu_e$ ($\overline{\psi}\overline{\psi}\to\psi\nu_e$) and the standard pair-annihilation $\psi\overline{\psi}\to \nu_\alpha \nu_\beta$, 
depending on the magnitude of the neutrino Yukawa coupling $(y_\nu)_{1e}$.
The Feynman diagrams for the semi-annihilation are shown in Fig.~\ref{fig:semi-ann}, and the cross section in the non-relativistic limit is calculated as
\begin{align}
\sigma_{\psi\psi\to\overline{\psi}\nu_e}v_\mathrm{rel}=\sigma_{\overline{\psi}\overline{\psi}\to\psi\nu_e}v_\mathrm{rel}=
\frac{81(y_{\nu})_{1e}^2(y_L)_{11}^{2}\sin^22\alpha~m_\psi^2(m_\varphi^2-m_\psi^2)^2}{1024\pi (4m_\psi^2+m_\psi^2v_\mathrm{rel}^2-m_\varphi^2)^2(m_\psi^2+2m_\varphi^2)^2},
\label{eq:semi}
\end{align}
where $v_\mathrm{rel}$ is the dark matter relative velocity, which is retained to properly describe the resonance region $m_\varphi\simeq 2m_\psi$.
Note that only the neutrino flavor $\alpha=e$ is possible since the neutrino Yukawa coupling $y_{\nu}$ is taken to be diagonal in our setup.
The cross section for the standard pair-annihilation shown in Fig.~\ref{fig:ann} is calculated as 
\begin{align}
\sigma_{\psi\overline{\psi}\to \nu_e\nu_e}{v}_\mathrm{rel}=
\frac{(y_{\nu})_{1e}^4 m_\psi^2}{16\pi}\left(\frac{\sin^2\alpha}{m_\psi^2+m_\varphi^2}+\frac{\cos^2\alpha}{m_\psi^2+m_{\tilde{\varphi}}^2}\right)^2,
\end{align}
The above formulae have been numerically validated using \texttt{CalcHEP}~\cite{Belyaev:2012qa}.

We use \texttt{FeynRules}~\cite{Alloul:2013bka} to generate the model files and \texttt{micrOMEGAs}~\cite{Alguero:2023zol} for the numerical calculations. 
In cases where the light mediator mass is close to the resonance ($m_\varphi\approx 2m_\psi$), the numerical calculations with \texttt{micrOMEGAs} may suffer from numerical instabilities, 
and the independent numerical package, \texttt{DRAKE}, provides more reliable results~\cite{Binder:2021bmg}.

The combination of the couplings $(y_{\nu})_{1e}\sin2\alpha$ is determined by the neutrino mass formula in Eq.~(\ref{eq:numass}). 
It follows that one finds the Yukawa coupling scales as $(y_L)_{11}\propto \left\{(y_{\nu})_{1e}\sin2\alpha\right\}^{-2}$ for $\sin2\alpha\ll1$, 
and therefore the semi-annihilation cross section in Eq.~(\ref{eq:semi}) is proportional to $\left\{(y_{\nu})_{1e}\sin2\alpha\right\}^{-2}$.

The numerical results are shown in Fig.~\ref{fig:relic}, where the mass ratio between the $\mathbb{Z}_3$ charged scalar and dark matter is fixed to $m_\varphi/m_\psi=1.9,1.99$ and $1.999$, 
while the mixing angle is also fixed to $\sin\alpha=0.01$ (left) and $0.1$ (right). 
The $\varphi$ mass is chosen to be close to the resonance $m_{\varphi}\approx 2m_\psi$ so that the semi-annihilation cross section is enhanced through the last diagram in Fig.~\ref{fig:semi-ann}. 
The color indicates the magnitude of $|(y_L)_{11}|$ as shown. 
In the red region, the relic abundance is dominated by the semi-annihilation process $\psi\psi\to\overline{\psi}\nu_e$, whereas the standard pair-annihilation $\psi\overline{\psi}\to\nu_e\nu_e$ dominates in the blue region. 
The gray region at the top and bottom are excluded by the perturbativity for the Yukawa couplings $(y_{\nu})_{1e}$ and $|(y_L)_{11}|$, respectively.

\subsection{Boosted dark matter}

If the semi-annihilation $\psi\psi\to\overline{\psi}\nu_e$ ($\overline{\psi}\overline{\psi}\to\psi\nu_e$) occurs in the Galaxy, a boosted (anti-)dark matter particle is produced. 
Such boosted dark matter may be detected on Earth through dark matter direct detection experiments and large-volume neutrino detectors.

The expected number of signal events is schematically given by $N_\mathrm{event}\sim \Phi_\mathrm{BDM} \sigma_p N_p t_\mathrm{exp}$ where 
$\Phi_\mathrm{BDM}$ is the boosted dark matter flux, $\sigma_p$ is the dark matter-proton scattering cross section, 
$N_p$ is the number of protons in the target detector and $t_\mathrm{exp}$ is the exposure time. 
The boosted dark matter flux $\Phi_\mathrm{BDM}$ is proportional to the semi-annihilation cross section in our model. 
Model-independent constraints were obtained in the previous work~\cite{BetancourtKamenetskaia:2025noa} where the semi-annihilation cross section is fixed at $10^{-26}~\mathrm{cm^3/s}$, 
assuming a fully asymmetic dark matter scenario, in which no anti-dark matter particles remain in the present-day Galaxy.
To apply the model-independent results in our case, we rescale the bound in the previous work with the factor 
$(2\times10^{-26}~\mathrm{cm^3s^{-1}})/\langle\sigma_{\psi\psi\to\overline{\psi}\nu_e}v_\mathrm{rel}\rangle$ where the factor of two arises because 
the observed relic abundance is composed of equal numbers of dark matter ($\psi$) and anti-dark matter ($\overline{\psi}$) particles in our case.\footnote{Strictly speaking, the boosted dark matter flux itself may depend on dark matter (semi-)annihilation cross section through the depletion of the central dark matter cusp close to the galactic center. However, the dependence turns out to be weak in our interesting parameter region~\cite{BetancourtKamenetskaia:2025ivl} and will be neglected in the following discussion.}

\begin{figure}[t]
 \begin{center}
 \includegraphics[width=4cm]{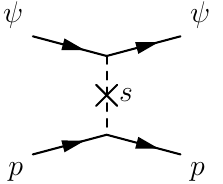}
\caption{Feynman diagram for boosted dark matter detection.}
\label{fig:dd}
 \end{center}
\end{figure}

The spin-independent scattering cross section with a proton $\sigma_p$ is induced by the light mediator $s$, as shown in Fig.~\ref{fig:dd}, and is given by 
\begin{align}
\sigma_{p}=\frac{f_p^2 \sin^22\theta m_\psi^4 m_p^4}{4\pi v^2 v_\sigma^2 m_s^4 (m_\psi+m_p)^2},
\end{align}
where $f_p\approx0.3$ is the effective coupling to a proton and $m_p=938~\mathrm{MeV}$ is the proton mass. 
In deriving the above expression, the momentum transfer has been neglected which allows us to directly apply the model-independent bound in the previous work~\cite{BetancourtKamenetskaia:2025noa}. 
If the momentum transfer is taken into account, the scattering cross section is expected to be slightly reduced for $m_s\lesssim 100~\mathrm{MeV}$ 
because the mediator propagator is suppressed by the finite momentum transfer. 

\begin{figure}[t]
\begin{center}
\includegraphics[width=7.5cm]{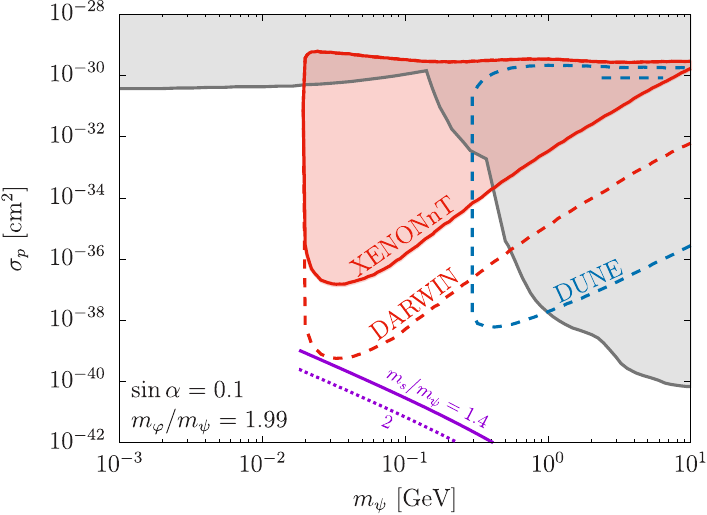}
\includegraphics[width=7.5cm]{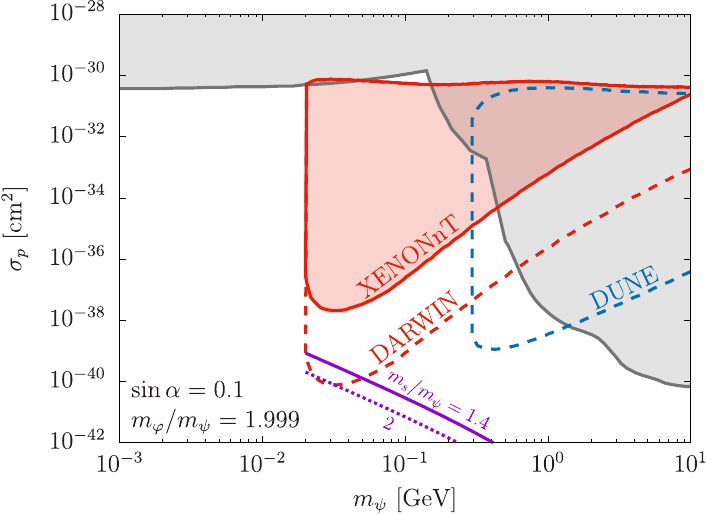}\\
\includegraphics[width=7.5cm]{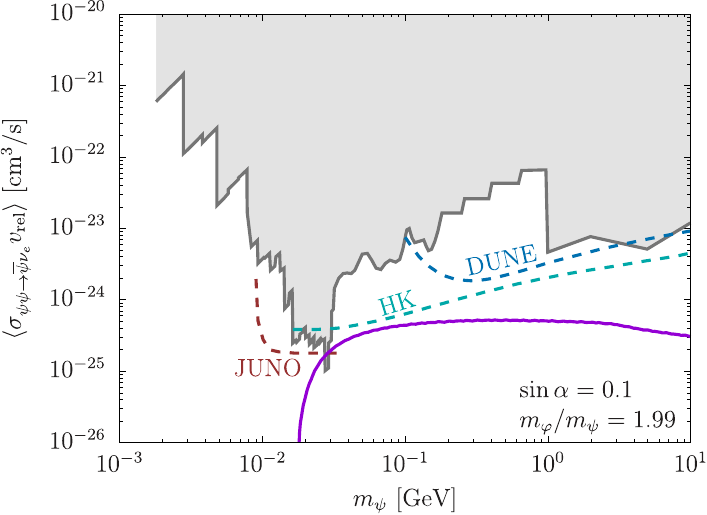}
\includegraphics[width=7.5cm]{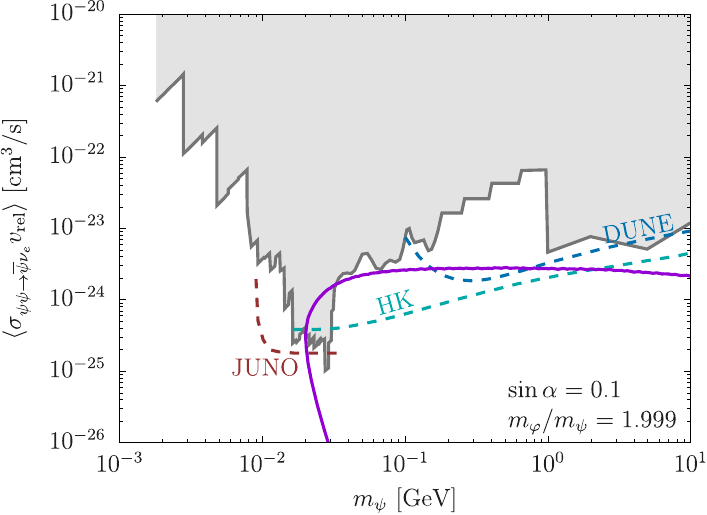}
\caption{
Experimental bound and detection prospects for boosted dark matter (top) and monochromatic neutrinos (bottom) produced by the semi-annihilation at the Galactic center where 
the mixing angle $\sin\alpha=0.1$, the $\mathbb{Z}_3$ charged mediator mass $m_\varphi$ is set to be $m_\varphi/m_\psi=1.99$ (left) and $1.999$ (right). 
The gray region is excluded by the current experiments while the dashed lines represent the future sensitivities (the detailed informaton for experiments are in the main texts). 
The solid and dashed purple lines reproduce the observed relic abundance $\Omega_\mathrm{DM} h^2=0.12$ for the light mediator mass $m_s/m_\psi=1.4$ and $2$, respevtively. 
}
\label{fig:bdm}
\end{center}
\end{figure}

The top panels in Fig.~\ref{fig:bdm} show the current bounds and the future sensitivities for the boosted dark matter where the $\mathbb{Z}_3$ charged mediator mass is fixed to be $m_\varphi/m_\psi=1.99$ on the left 
and $1.999$ on the right. 
We also fix the Higgs mixing angle $\sin\theta=10^{-4}$ as in Fig.~\ref{fig:relic} for the relic abundance calculation. 
The purple solid and dotted lines show the predicted scattering cross section obtained for parameters sets reproducing the observed relic abundance for the fixed light mediator mass $m_s/m_\psi=1.4$ and $2$, respectively.
In the dark matter mass region $m_\psi\lesssim20~\mathrm{MeV}$, the annihilation cross section is too large to reproduce the observed relic abundance. Therefore, the purple curves terminate at the value.
The gray region is excluded by conventional direct detection experiments under the assumption of a non-relativistic Galactic dark matter halo, CRESST~\cite{CRESST:2015txj, CRESST:2017ues}, cosmic ray boosted dark matter~\cite{Bringmann:2018cvk}, 
CMB observations~\cite{Xu:2018efh}, Lyman-$\alpha$~\cite{Rogers:2021byl} and gas cloud cooling~\cite{Bhoonah:2018wmw}. 
The red region is excluded by the direct detection experiment XENONnT~\cite{XENON:2023cxc} for boosted dark matter 
produced through semi-annihilation in the Galactic center~\cite{BetancourtKamenetskaia:2025noa}. 
The regions enclosed by the red and blue dashed lines represent the future sensitivities of DARWIN~\cite{DARWIN:2016hyl} and DUNE~\cite{DUNE:2015lol}.\footnote{For the DARWIN sensitivity, we simply scale the current XENONnT limit by a factor $270$ since both experiments use the same target nucleus at the same location. For the DUNE sensitivity, we assume a $40$ kton detector and 10 years of exposure with an energy range of $50~\mathrm{MeV}-10~\mathrm{GeV}$. Further details can be found in Ref.~\cite{BetancourtKamenetskaia:2025noa}.}
The upper parts of the XENONnT exclusion bound and the projected DARWIN and DUNE sensitivities at $\mathcal{O}(10^{-30})~\mathrm{cm^2}$ are truncated due to attenuation in the overburden, 
since the scattering cross section is so large that the boosted dark matter loses its energy before reaching the underground detectors.
We find that the semi-annihilation boosted dark matter can be detected through DARWIN if the dark matter mass is in the range of 
$20~\mathrm{MeV}\lesssim m_\psi\lesssim 50~\mathrm{MeV}$ for $m_s/m_\psi=1.4$ and a $\mathcal{O}(0.1)\%$ resonance tuning. 

We briefly comment on dark matter-electron scattering. 
Our Dirac fermion dark matter candidate is leptophilic in the original $\mathbb{Z}_3$ symmetric model~\cite{Ma:2007gq, Aoki:2014cja}, 
it is natural to ask whether it can also be detected through scattering off electrons instead of protons. 
The scattering cross section with an electron in the non-relativistic regime is mediated by the $u$-channel exchange of the charged scalar $\eta^\pm$, and is given by
\begin{align}
\sigma_e=\frac{(y_{\nu})_{e1}^4 m_\psi^2m_e^2}{16\pi m_{\eta}^4(m_\psi+m_e)^2}.
\end{align}
This cross section is far below the current experimental sensitivity~\cite{XENON:2019gfn, DAMIC:2019dcn, SENSEI:2020dpa, PandaX-II:2021nsg, DarkSide:2022knj} because the charged scalar mass must be above the TeV scale 
to satisfy the constraints of the electroweak precision tests.

\subsection{Monomchromatic neutrino signals}
The bottom panels in Fig.~\ref{fig:bdm} show the current bounds and future sensitivities for monochromatic neutrinos from the Galactic center 
produced through semi-annihilation where the mass ratio is fixed to $m_\varphi/m_\psi=1.99$ on the left and $1.999$ on the right. 
The purple solid line shows the semi-annihilation cross section in the current Galaxy for parameters sets reproducing the observed relic abundance. 
The semi-annihilation cross section is significantly enhanced by the resonance. 
In contrast to the boosted dark matter signal, the monochromatic neutrino signal is independent of the light mediator mass $m_s$. 
The gray region is excluded by Borexino~\cite{Borexino:2010zht, Borexino:2019wln}, KamLAND~\cite{KamLAND:2011bnd} and Super-Kamiokande~\cite{Super-Kamiokande:2015qek, Super-Kamiokande:2020sgt, Olivares-DelCampo:2017feq}, 
as summarized in Ref.~\cite{Arguelles:2019ouk}. 
The brown, light green and blue dashed lines represent the future sensitivities of JUNO~\cite{JUNO:2015zny, Arguelles:2019ouk}, 
Hyper-Kamiokande~\cite{Hyper-Kamiokande:2018ofw, Bell:2020rkw, RamosCascon:2025gso} and DUNE~\cite{DUNE:2015lol, DUNE:2020lwj} (See also IceCube~\cite{IceCube:2025sev}). 
The bounds and sensitivities have been rescaled by a factor four in the above references to adapt them for our case.
The factor $4$ originates from the fact that in our case the dark matter particle is a Dirac fermion and only one neutrino is produced per semi-annihilation $\psi\psi\to\overline{\psi}\nu_e$.

The figures show that the dark matter mass region of $30~\mathrm{MeV}\lesssim m_\psi\lesssim 2~\mathrm{GeV}$ with $m_\varphi/m_\psi=1.999$ 
falls within the projected sensitivities of Hyper-Kamiokande~\cite{Hyper-Kamiokande:2018ofw, Bell:2020rkw} and DUNE~\cite{DUNE:2015lol, DUNE:2020lwj} 
through monochromatic neutrinos from the Galactic center. 
Combining this result with the boosted dark matter search, we find that dark matter with $30~\mathrm{MeV}\lesssim m_\psi\lesssim40~\mathrm{MeV}$ and $m_s/m_\psi=1.4$ 
can be simultaneously probed by both searches, provided that the $\mathbb{Z}_3$-charged scalar mass is tuned at the $\mathcal{O}(0.1)~\%$ level.

The standard annihilation process $\psi\overline{\psi}\to\nu_e\nu_e$ also produces monochromatic neutrinos through an $s$-wave. 
However, unlike the semi-annihilation channel, its annihilation cross section is not resonantly enhanced. 
Consequently, the predicted signal remains well below the projected experimental sensitivities.

\section{Conclusions}
\label{sec:conc}

Once the SM is extended with non-minimal dark sectors, non-standard annihilation processes such as semi-annihilations, $3\to2$ and $4\to2$ SIMP processes, 
and dark matter conversion processes can occur. 
As a result, dark matter particles with some momentum can be produced in the final states. 
Such boosted dark matter can be an interesting signature to explore non-minimal dark sectors. 
In particular, the detectability for the boosted dark matter originating from the semi-annihilation into a pair of an anti-dark matter particle and a neutrino was studied in the previous work in a model independent manner. 
However, it is non-trivial how one can construct a concrete model with such a specific semi-annihilation channel.

In this work, we have constructed a concrete model realizing boosted dark matter from the semi-annihilation $\psi\psi\to\overline{\psi}\nu_e$ ($\overline{\psi}\overline{\psi}\to\psi\nu_e$). 
The original $\mathbb{Z}_3$ symmetric radiative neutrino mass model at the two-loop level was extended with a global U(1)$_L$ symmetry. 
After the spontaneous symmetry breaking, the $\mathbb{Z}_3$ symmetry remains, which forbids tree level neutrino masses and guarantees stability of the dark matter candidate at the same time.
Thus dark matter and neutrino phenomenology are closely correlated with each other in our model.
In particular, the semi-annihilation $\psi\psi\to\overline{\psi}\nu_e$ naturally arises. 
We imposed the relevant experimental and theoretical constraints such as cLFV, the electroweak precision tests, the Higgs mixing angle, and potential stability. 
The dark matter relic abundance is mainly determined by the semi-annihilation and the standard annihilation $\psi\overline{\psi}\to \nu_e\nu_e$, depending on the parameter space. 
In the sub-GeV mass range, the thermal relic abundance is predominantly determined by the semi-annihilation process. At the same time, this process generates an observable boosted dark matter flux, 
leading to relativistic proton scattering in direct detection experiments. 
In addition, the monochromatic neutrino accompanied by the boosted dark matter can also be detected in some parameter space. 
Combining the projected sensitivities for the boosted dark matter and monochromatic neutrino, we found that these are simultaneously detectable if the dark matter mass is in the range of $30-40~\mathrm{MeV}$ with a $\mathcal{O}(0.1)~\%$ tuning. 
Such simultaneous detections in both experiments within this dark matter mass range would provide a distinctive signature of our model.

\section*{Acknowledgments}
This work was supported in part by JSPS Grant-in-Aid for Scientific Research KAKENHI
Grants No. 25K23378 (M.F.), 25H02179 (T.T.) and 25K07279 (T.T.).


\bibliographystyle{utphysmod}
\bibliography{references}


\end{document}